\documentclass[superscriptaddress, reprint,nofootinbib, amsmath,amssymb, floatfix]{revtex4-1}

\usepackage[  ]{graphicx} 
\usepackage{dcolumn} 
\usepackage{bm,subfigure} 
 
\newcommand{\point}{\raise0.7ex\hbox{.}}

\usepackage{threeparttable}

\begin{document}

\preprint{APS/123-QED}

\title{Stimulated Brillouin scattering enhancement in silicon inverse opal waveguides}

\author{M. J. A. Smith}
\email{m.smith@physics.usyd.edu.au}
\affiliation{Centre for Ultrahigh bandwidth Devices for Optical Systems (CUDOS) and Institute of Photonics and Optical Science (IPOS), School of Physics, The University of Sydney, NSW 2006, Australia}
 \affiliation{Centre for Ultrahigh bandwidth Devices for Optical Systems (CUDOS), School of Mathematical and Physical Sciences, University of Technology Sydney, NSW 2007, Australia }
\author{C. Wolff}
 \affiliation{Centre for Ultrahigh bandwidth Devices for Optical Systems (CUDOS), School of Mathematical and Physical Sciences, University of Technology Sydney, NSW 2007, Australia }
\author{C. Martijn de Sterke}
\affiliation{Centre for Ultrahigh bandwidth Devices for Optical Systems (CUDOS) and Institute of Photonics and Optical Science (IPOS), School of Physics, The University of Sydney, NSW 2006, Australia}
\author{M. Lapine}
 \affiliation{Centre for Ultrahigh bandwidth Devices for Optical Systems (CUDOS), School of Mathematical and Physical Sciences, University of Technology Sydney, NSW 2007, Australia }
\author{B. T. Kuhlmey}
\affiliation{Centre for Ultrahigh bandwidth Devices for Optical Systems (CUDOS) and Institute of Photonics and Optical Science (IPOS), School of Physics, The University of Sydney, NSW 2006, Australia}
\author{C. G. Poulton}
 \affiliation{Centre for Ultrahigh bandwidth Devices for Optical Systems (CUDOS), School of Mathematical and Physical Sciences, University of Technology Sydney, NSW 2007, Australia }

\date{\today} 

\begin{abstract} \noindent
	Silicon is an ideal material for on-chip applications, however its poor acoustic properties limit its performance for important optoacoustic applications, particularly for Stimulated Brillouin Scattering (SBS). We theoretically show that silicon inverse opals exhibit a strongly improved acoustic performance that enhances the bulk SBS gain coefficient by more than two orders of magnitude. We also design a waveguide that incorporates silicon inverse opals and which has SBS gain values that are comparable with chalcogenide glass waveguides. This research opens new directions for opto-acoustic applications in on-chip material systems.  
\end{abstract}

\maketitle

   \section{Introduction} In recent years, considerable attention has been directed at enhancing stimulated Brillouin scattering (SBS) in silicon waveguides for on-chip applications \cite{rakich2012giant,shin2013tailorable,wolff2014germanium,van2015interaction,kittlaus2015large,wolff2016brillouin,sarabalis2016guided}. However, the intrinsically poor properties of silicon has made photonic integration of SBS-based devices, such as Brillouin lasers and compact microwave signal processors, a fundamentally difficult task \cite{eggleton2013inducing}. The poor performance of silicon for SBS applications can be attributed to a combination of high mechanical stiffness, low photoelasticity, and high acoustic losses, relative to other materials such as chalcogenide and silica glasses \cite{pant2011chip}. However, the existence of highly-refined fabrication facilities for silicon, together with the potential for integration with other photonic components, makes silicon an extremely attractive platform for the cost-effective mass-production of SBS-based devices \cite{eggleton2013inducing}.

Existing work on the enhancement of SBS in silicon has primarily focused on waveguide designs that exploit radiation pressure (boundary motion) contributions to improve the electrostrictive response of the structure, treating the bulk properties of silicon as fixed. Here we propose a new method to enhance the intrinsic properties of silicon by using a metamaterial design based on silicon inverse opals. This involves structuring silicon with a lattice of interpenetrating spherical air holes (see Fig. \ref{fig:1}a). The choice of air as the structuring material is physically motivated by its higher compressibility, lower speed of sound, and considerably lower acoustic losses, relative to most bulk dielectric materials. The principal outcome is that silicon inverse opals are less mechanically stiff than conventional silicon, which reduces the acoustic speed of sound in the material, enabling both light and sound to be confined within the same region. In addition, the reduction of the material quantity of silicon reduces the acoustic losses. The air-structuring of silicon thereby overcomes the fundamental constraints of silicon and gives it a clear advantage for SBS applications.

To the best of our knowledge, this is the first study on the bulk SBS properties of air-structured metamaterials, complementing recent work on the control of SBS \cite{Tchahame2016surface,florez2016brillouin} in hollow-core silica fibres and recent theoretical work by the authors where metamaterials were shown to give enhancement and suppression of the bulk SBS \cite{smith2016control,smith2016stimulated}. Here we report the extension of \cite{smith2016control,smith2016stimulated} to interconnected air inclusions and extend theoretical methods to evaluate all materials constants and wave propagation parameters present in the bulk SBS gain coefficient.

Having characterised the bulk properties of silicon inverse opals, we then numerically demonstrate the usefulness of the material with a waveguide design (see Fig.~\ref{fig:1}b). A fundamental issue with silicon waveguides is that whilst optical confinement is in principle straightforward, since the refractive index of silicon is high, acoustic mode confinement is difficult. The latter is due to the high speed of sound in silicon \cite{rakich2012giant,eggleton2013inducing} and the low speed of sound in common substrates such as fused silica glass \cite{abedin2005observation}. The poor acoustic performance of silicon drives a poor acousto-optic overlap and subsequently weak SBS in a waveguide. Here we demonstrate that both a high refractive index and a low acoustic velocity can be achieved in bulk silicon inverse opals, which gives rise to much stronger acousto-optic overlap in silicon inverse opal waveguides. The improved acousto-optic overlap enhances the SBS gain in the waveguide to values comparable with uniform chalcogenide waveguides \cite{pant2011chip,wolff2014germanium} and germanium waveguides \cite{wolff2014germanium}, and therefore has promising implications for silicon-based SBS devices.

\section{Bulk SBS gain for a structured silicon} \label{sec:sbsbulk} 

\begin{figure}[t] \centering 
	\includegraphics[width=0.39\linewidth]{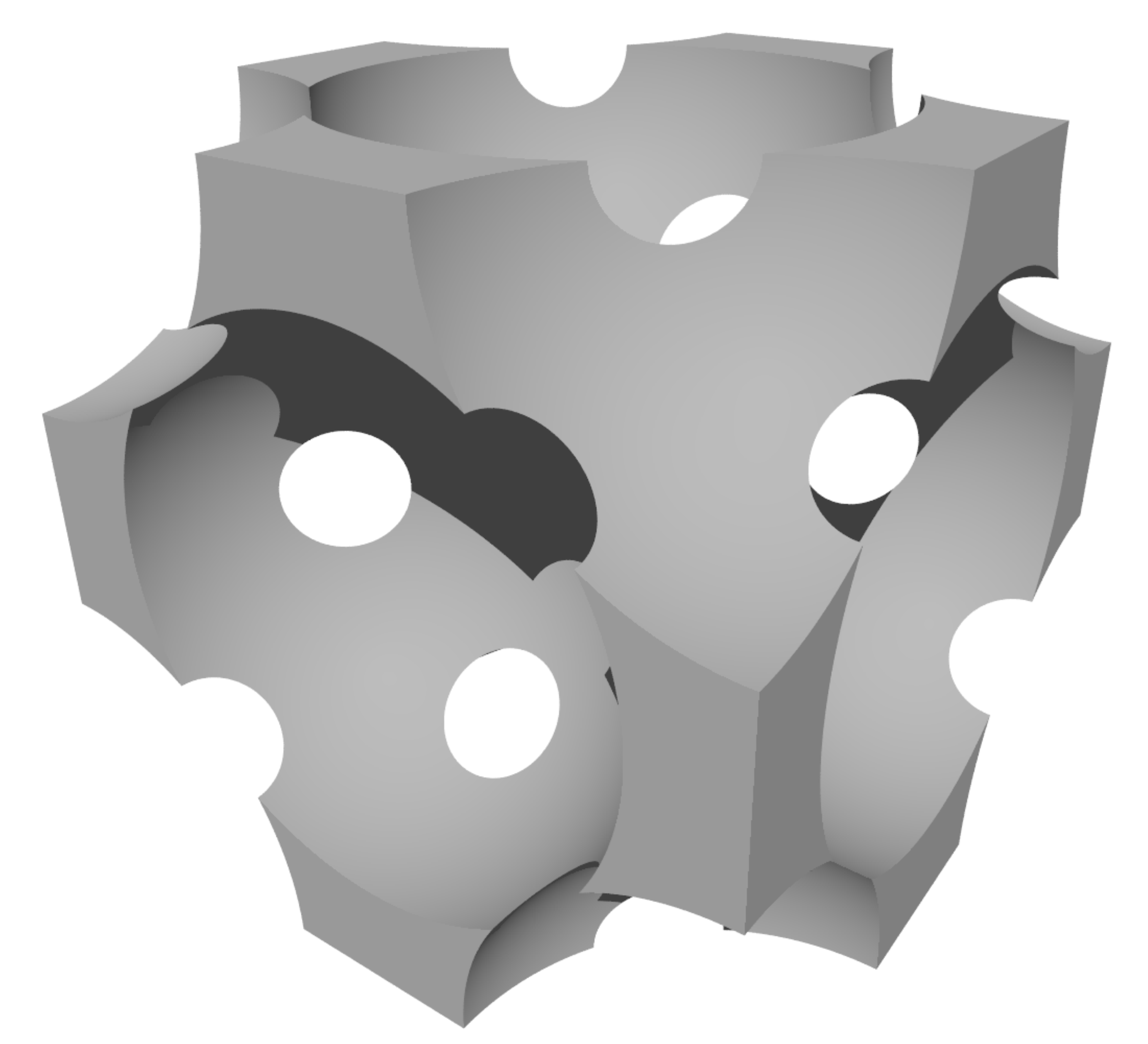} \centering 
	\includegraphics[width=0.49\linewidth]{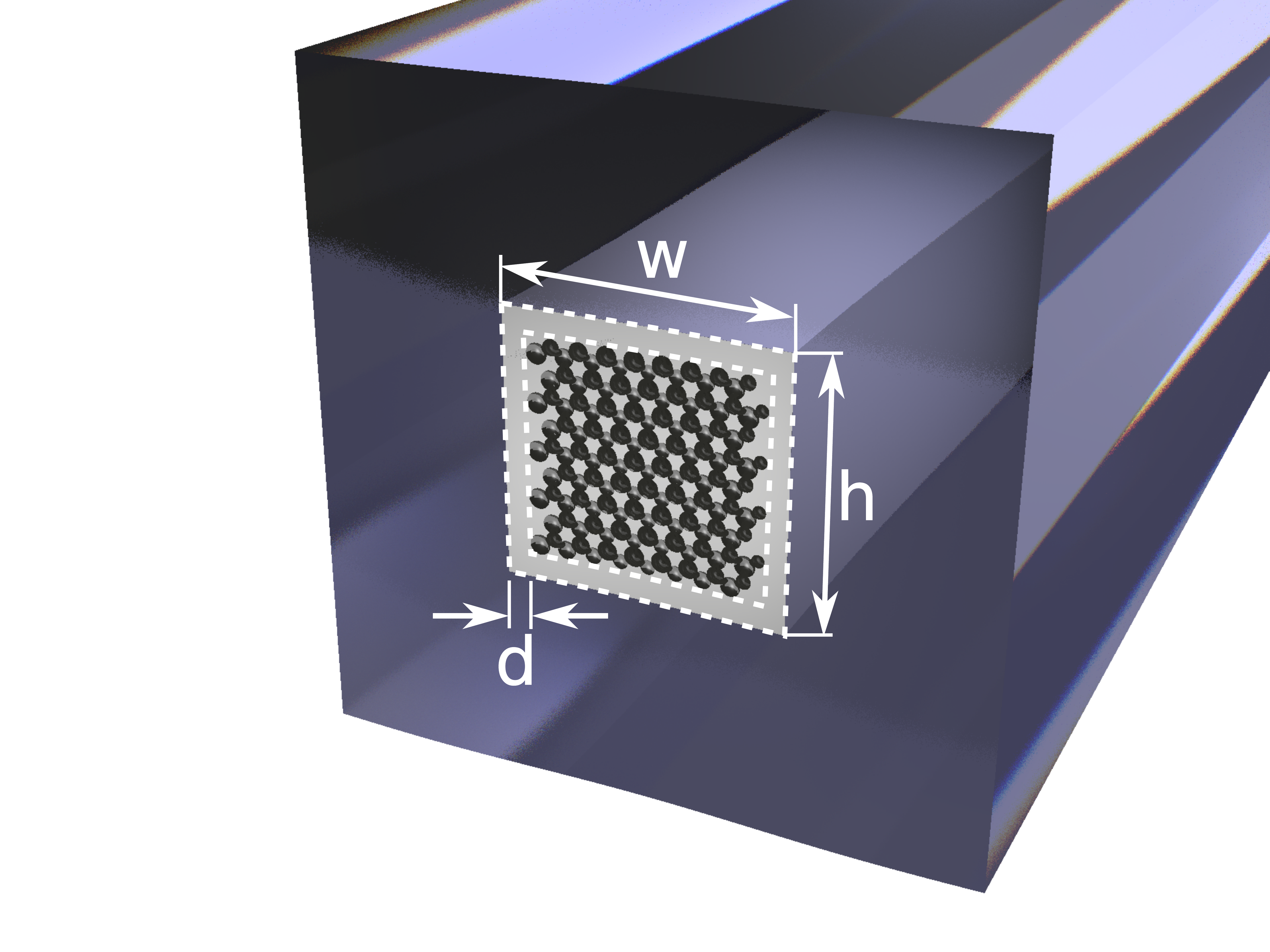} \caption{(a) Fundamental unit cell for face-centred cubic (FCC) lattice of interpenetrating air holes in silicon. (b) Waveguide design comprising silicon inverse opal core with silicon cladding (width $d$) embedded in fused silica substrate where $w=550~\mathrm{nm}$ and $h = 500~\mathrm{nm}$. \label{fig:1}} 
\end{figure}

We first compute the SBS gain of a nanoscale-structured metamaterial comprising a face-centered cubic (FCC) lattice of interpenetrating air holes in silicon $\langle 100 \rangle$ (see Fig. \ref{fig:1}a). The period is chosen to ensure the metamaterial is both optically and acoustically subwavelength at an incident wavelength of $\lambda_1 = 1550 \, \mathrm{nm}$ and we assume negligible optical losses.  Numerical results presented here are obtained using COMSOL 4.4 \& 5.1. The backwards SBS power gain coefficient for an incident optical plane wave propagating through a cubic or isotropic bulk material is \cite{powers2011fundamentals,kobyakov2010stimulated} 
\begin{equation}
	\label{eq:gp} g_P = \frac{4 \pi^2 \gamma_\mathrm{12}^{\; 2}}{ n c \lambda_1^2 \rho V_\mathrm{A} \Gamma_\mathrm{B}}, 
\end{equation}
where $\gamma _\mathrm{12} = p_{12} n^4$ is the electrostrictive stress induced by the incident optical field, $p_{12}$ is the photoelastic constant (an element of the photoelastic tensor which applies for all plane wave polarisations), $n$ is the refractive index, $c$ is the speed of light in vacuum, $\rho$ is the mass density of the medium, $V_\mathrm{A}$ is the longitudinal acoustic velocity, and $\Gamma_\mathrm{B}$ is the Brillouin line width.

We characterise the SBS properties of silicon inverse opals by calculating effective parameters for all optical, acoustic, and opto-acoustic terms present in \eqref{eq:gp} \cite{smith2016control,smith2016stimulated}. The photoelastic constant $p_{12}$ is determined by mechanically perturbing the boundary of the unit cell and calculating the corresponding change in the effective inverse permittivity tensor. The deformed unit cell is obtained by solving the acoustic wave equation \cite{auld1973acoustic} under a static boundary loading with continuity conditions imposed at all mechanical interfaces. Unlike in previous work \cite{smith2016control,smith2016stimulated}, for a silicon inverse opal we must impose free-edge boundary conditions $\sigma_{ij} \hat{n}_j \big|_{ 
\partial I} = 0$ (solid-vacuum conditions) at all air interfaces, where $\sigma_{ij}$ denotes the stress tensor, and $\hat{n}_j$ is the normal vector to the internal boundary $ 
\partial I$, in place of standard continuity conditions (solid-solid conditions). Whilst the calculation of the effective optical properties is unchanged from \cite{smith2016control,smith2016stimulated}, a subtle modification is required to calculate the acoustic parameters $V_\mathrm{A}$ and $\Gamma_\mathrm{B}$: for these terms we solve the eigenvalue problem for the acoustic wave equation \cite{auld1973acoustic} with free edge conditions on $ 
\partial I$, and acoustic Bloch conditions imposed only upon the remnants of the silicon background at the cell edge, since air gaps do not form part of the acoustic domain.

With these modifications, we characterise the bulk properties of silicon inverse opals with air filling fractions $f = 75\%$, $80\%$, and $85\%$. The range of filling fractions for an inverse opal is $74 \% \lesssim f \lesssim 96\%$ where the bounds are the sphere-touching limit and the point where the two different inverse opal sites disconnect (i.e., when the tetrahedral and octahedral interstitial sites are no longer connected \cite{gaillot2005photonic}). However, in practice this range is smaller because the material is not structurally stable at very high filling fractions. Here we impose an upper bound of $f =85\%$ which is chosen to ensure that the interstitial site widths are of comparable size. This bound is close to $f=88\%$ which is the greatest value achieved for micron-scale inverse opals \cite{blanco2000large}.

\begin{table*}
	[t] \centering \caption{\label{tab:table1} \bf Calculated bulk optical, acoustic, optoacoustic parameters, and SBS gain coefficient at $\lambda_1 = 1550~\mathrm{nm}$.$^a$ }
	
	{ \small 
	\begin{tabular}
		{lc c ccc cc ccc ccc cc cc cc ccc ccc cc} \hline $f$ & $a$ &$n$ & $\rho$ & $V_\mathrm{A}$ & $ {\Omega_\mathrm{B}}/{2\pi}$ & $ {\Gamma_\mathrm{B}}/{2\pi}$ & $C_{11}$ & $C_{12}$ & $C_{44}$ &$\eta_{11}$ & $\eta_{12}$ & $\eta_{44}$ & $p_{11}$ & $p_{12}$ & $p_{44}$ & $g_P$ \\
		\hline $ 75\%$ &17.754 &1.68 & 582.25 &5210 & 11.3&21.3 &15.81 & 7.96 & 6.8 & 0.42 & 0.34 & 0.042 &-0.10 & 0.18 & -0.14 & $1.7 \times 10^{-10}$ \\
		$ 80\%$ &18.1818 &1.53 & 465.80 &4591 & 9.1&9.6 & 9.82 & 5.55 & 4.12 & 0.18 & 0.14 & 0.018 &-0.09& 0.22 & -0.15 & $3.9 \times 10^{-10}$ \\
		$ 85\%$ &18.6646 &1.38 & 349.35 &3985 & 7.1&3.8 &5.55 & 3.57 & 2.15 & 0.07 & 0.06 & 0.007 &-0.05 & 0.23 & -0.14 & $8.2 \times 10^{-10}$ \\
		\hline 
	\end{tabular}
	\begin{tablenotes}
		 \item $^a$ Here $f$ is the air filling fraction for sphere radii $a$ and lattice period $d= 50$ (both in units of $\left[\mathrm{nm}\right]$), $n$ the refractive index, $\rho$ is the mass density (in $\left[ \mathrm{kg} \cdot \mathrm{m}^{-3}\right]$), $V_A$ the longitudinal acoustic velocity (in $\left[ \mathrm{m} \cdot \mathrm{s}^{-1} \right]$), $\Omega_\mathrm{B}/(2 \pi)$ the Brillouin frequency shift (in $\left[\mathrm{GHz}\right]$), $\Gamma_\mathrm{B} / (2\pi)$ the Brillouin linewidth (in $\left[\mathrm{MHz}\right]$), $C_{ij}$ are stiffness tensor coefficients (in $\left[ \mathrm{GPa} \right]$), and $\eta_{ij}$ are phonon viscosity tensor coefficients (in $\left[ \mathrm{mPa} \cdot \mathrm{s} \right]$). $p_{ij}$ are photoelastic tensor coefficients, and $g_{P}$ is the gain coefficient (units of $ \left[\mathrm{W}^{-1} \cdot \mathrm{m} \right]$). Subscripts are represented in Voigt notation. 
	\end{tablenotes}
	
	} 
\end{table*}

Table \ref{tab:table1} shows the calculated optical, acoustic, opto-acoustic, and SBS parameters as a function of filling fraction. As expected, the refractive index of the metamaterial decreases with increasing air content $f$. The photoelastic constant $p_{12}$ increases with increasing $f$ to values more than one order of magnitude greater than in pure silicon ($p_{12}^\mathrm{Si} = 0.017$). Despite large enhancements in the photoelastic constant, the electrostrictive stress constant $\gamma_{12} = n^4 p_{12}$ decreases with increasing $f$ because the drop in permittivity is much stronger. The gain coefficient increases with increasing filling fraction, due to positive contributions from all terms in \eqref{eq:gp} excluding $\gamma_{12}$. The increased gain can be understood with reference to the purely acoustic parameters in the table, which all decrease with increasing air filling fraction. A reduced Brillouin linewidth has the biggest overall contribution to the gain enhancement, dramatically falling from $\Gamma_\mathrm{B}^\mathrm{Si} / (2\pi) = 320 \, \mathrm{MHz}$ to $\Gamma_\mathrm{B} / (2\pi) = 3.8 \, \mathrm{MHz}$ at $f=85\%$, and dominating improvements in all other terms. Such behaviour was previously observed for silica spheres in silicon \cite{smith2016stimulated} at high filling fractions. The improved optical and acoustic performance of air-structured silicon ultimately offsets the reduced $\gamma_{12}$ to give a bulk SBS gain of $g_P = 8.2 \times 10^{-10} \, \mathrm{W}^{-1} \mathrm{m}$ at $f = 85\%$, more than two orders of magnitude above that of uniform silicon ($ g_P^\mathrm{Si} = 2.4 \times 10^{-12} \, \mathrm{W}^{-1} \mathrm{m} $).  We observe that bulk porous silicon with an air filling fraction of $f\gtrsim 75\%$ has a sufficiently low stiffness to support confined acoustic modes in a silica environment, and for $f\lesssim 80\%$ it has a sufficiently high permittivity to support guided optical modes in such an environment. This suggests that silicon inverse opal waveguides with $ 75\% \lesssim f \lesssim 80\%$ exhibit strong SBS. This is further investigated in Section~\ref{sec:SBSwaveguide}. 

The values for $p_{44}$ in Table~\ref{tab:table1} are obtained with the isotropic approximation $p_{44} = (p_{11} - p_{12})/2$, an approximation that is common in the optical, acoustic, and SBS literature. Also, we calculate $\eta_{11}$ directly using SBS parameters \cite{smith2016control}, and set $\eta_{44} \approx \eta_{11} /10$, which is consistent with experimental data for semiconductors \cite{helme1978phonon}. The remaining $\eta_{12}$ element is then obtained using the isotropic approximation $\eta_{12} = \eta_{11} - 2\eta_{44}$. We find that these approximations for $\eta_{12}$ and $\eta_{44}$ do not affect the results substantially since all components of the viscosity tensor have a small magnitude. The use of an isotropic $p_{44}$ slightly perturbs the gain values obtained for waveguides in Section~\ref{sec:SBSwaveguide}, however, we anticipate that values calculated for the SBS gain in a waveguide are of the same order of magnitude.

\section{SBS in a silicon waveguide} \label{sec:SBSwaveguide} Having characterised the bulk properties of silicon inverse opals, we now investigate SBS in silicon inverse opal waveguides surrounded by a silica substrate, as would be required due to the need in a physical device to mechanically support the waveguide. We evaluate the SBS gain in a waveguide using the representation  \cite{wolff2014stimulated}
\begin{equation}
	\label{eq:gainwg} g_P^\mathrm{wg} = \left( \frac{4 \pi c}{\lambda_1 | P | U_\mathrm{A}} \right) |\xi|^2 Q_\mathrm{A}, 
\end{equation}
where $P$ denotes the energy flux for the optical mode, $U_\mathrm{A}$ is the energy density of the acoustic mode, $\xi$ is the opto-acoustic overlap integral, and $Q_\mathrm{A}$ is the acoustic quality factor of the waveguide, and the brackets contain normalisation factors.

In principle, a simple design comprising a rod of porous silicon embedded in silica confines both sound and light and can therefore support SBS. However, the low dielectric contrast and relatively low stiffness of silicon inverse opals means that the SBS-gain of such structures is low; the optical and acoustic modes extend well into the substrate, resulting in poor opto-acoustic overlap. The waveguide we propose is shown in Fig.~\ref{fig:1}b, and comprises a silicon inverse opal core surrounded by a silicon cladding layer that is suspended in fused silica. The thickness of the silicon shell layer plays an important role in the predicted SBS gain, and here we investigate optimal layer widths surrounding an inverse opal core. Introducing an additional layer of unstructured silicon around a core of porous silicon (see Fig.~\ref{fig:1}b) improves both the optical and the acoustic confinement through two distinct mechanisms. Optical confinement is improved by the high permittivity of the cladding, as the silicon walls bear some similarity to a slot waveguide. Acoustic confinement is improved by the high stiffness of the silicon shell which reduces the tails of the acoustic mode in the silica cladding. This reduction of the optical and acoustic mode areas enhances the overall SBS-gain of our waveguide to levels competitive with experimentally demonstrated SBS platforms.

\begin{figure}
	[t] \centering 
	\includegraphics[width=0.49\linewidth]{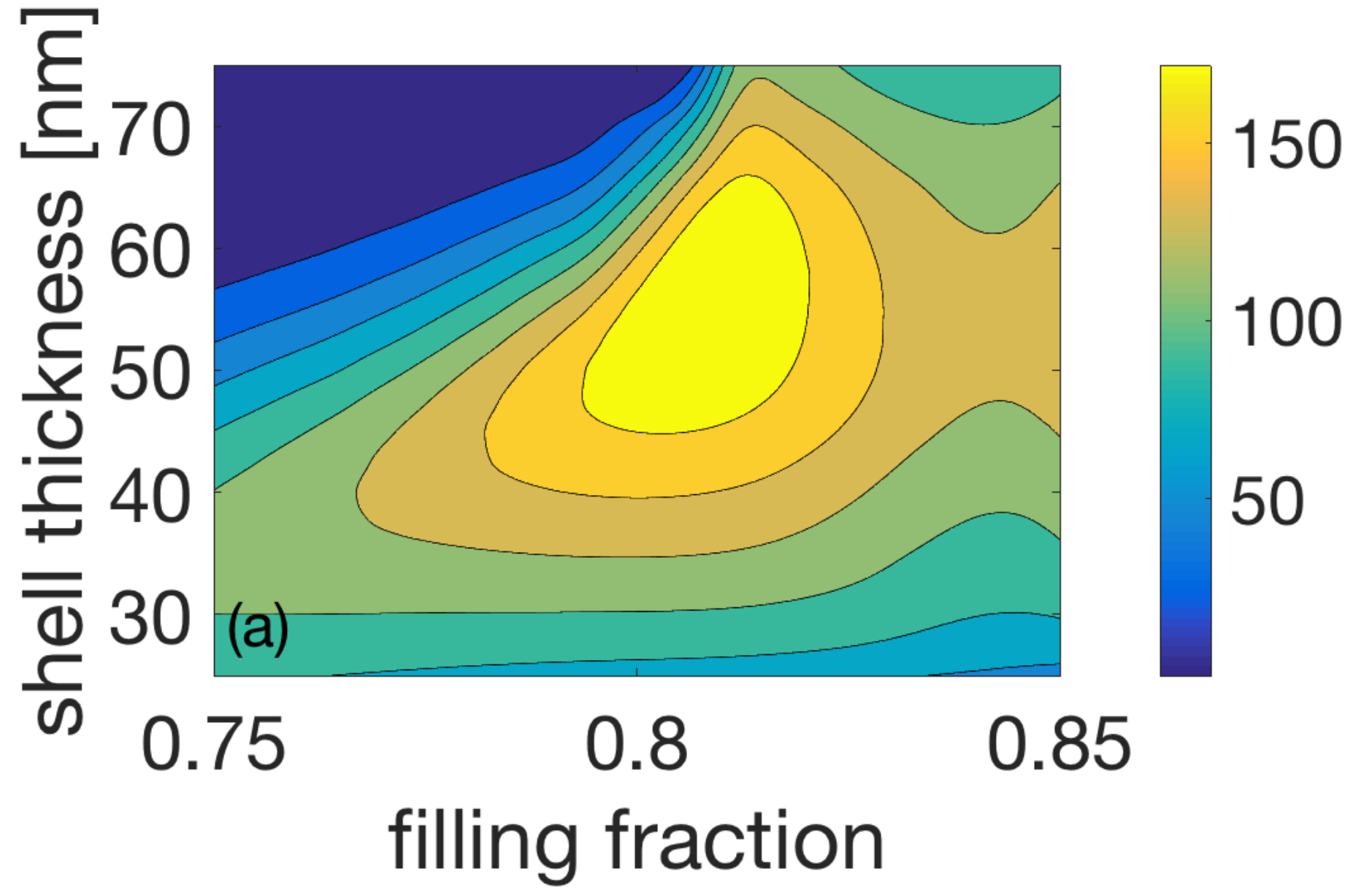} 
	\includegraphics[width=0.49\linewidth]{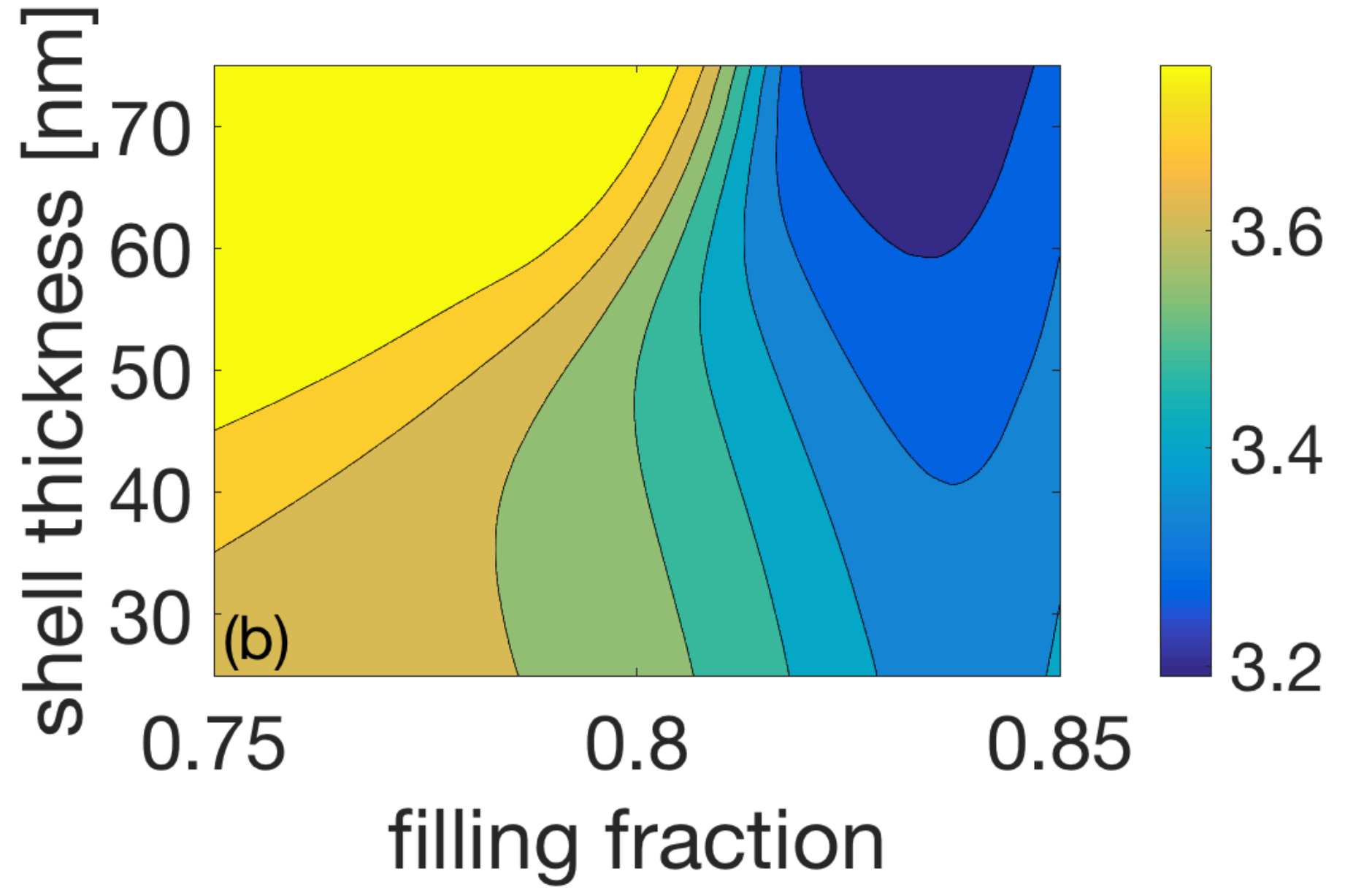} 
	\includegraphics[width=0.49\linewidth]{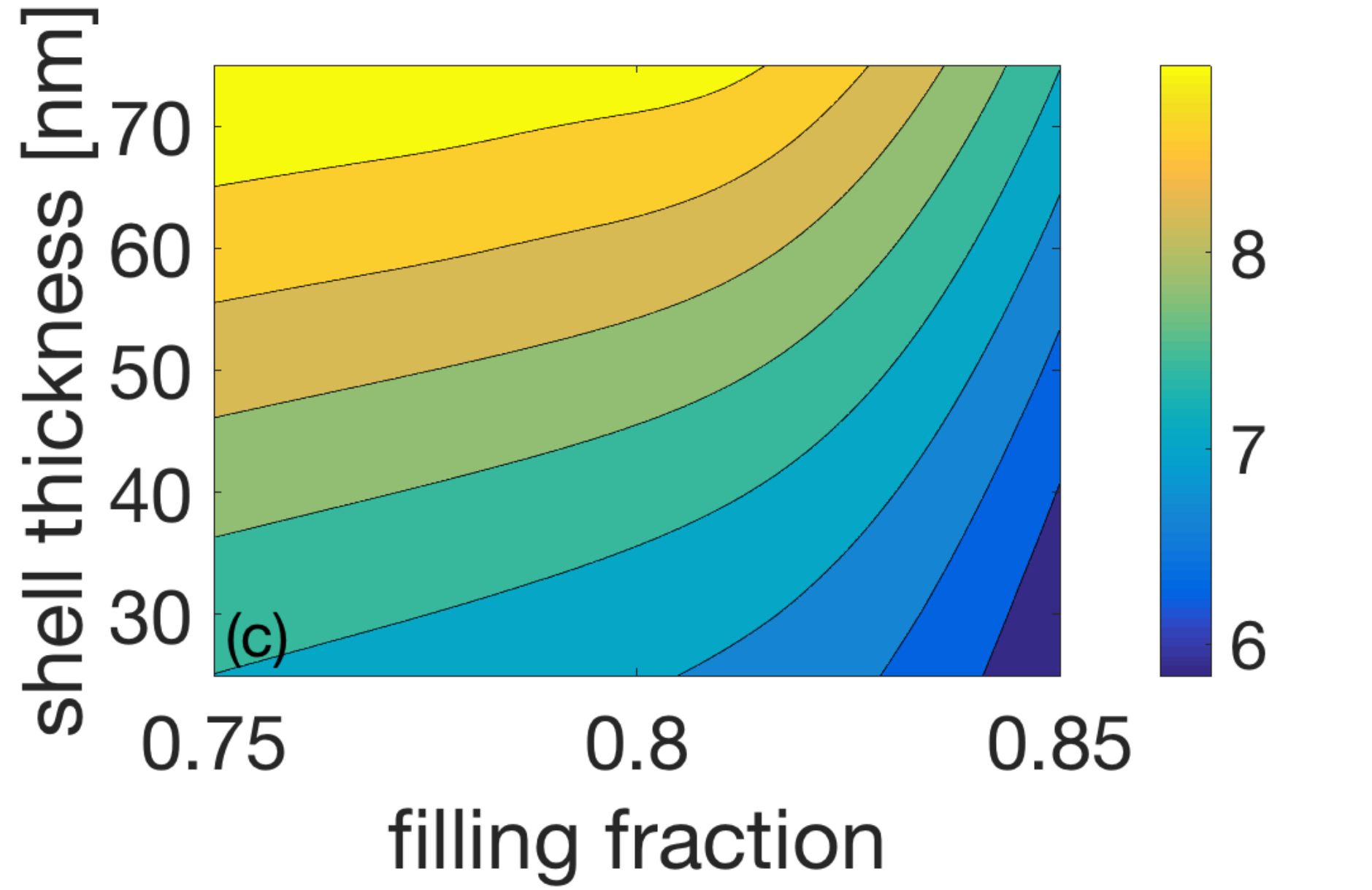} 
	\includegraphics[width=0.49\linewidth]{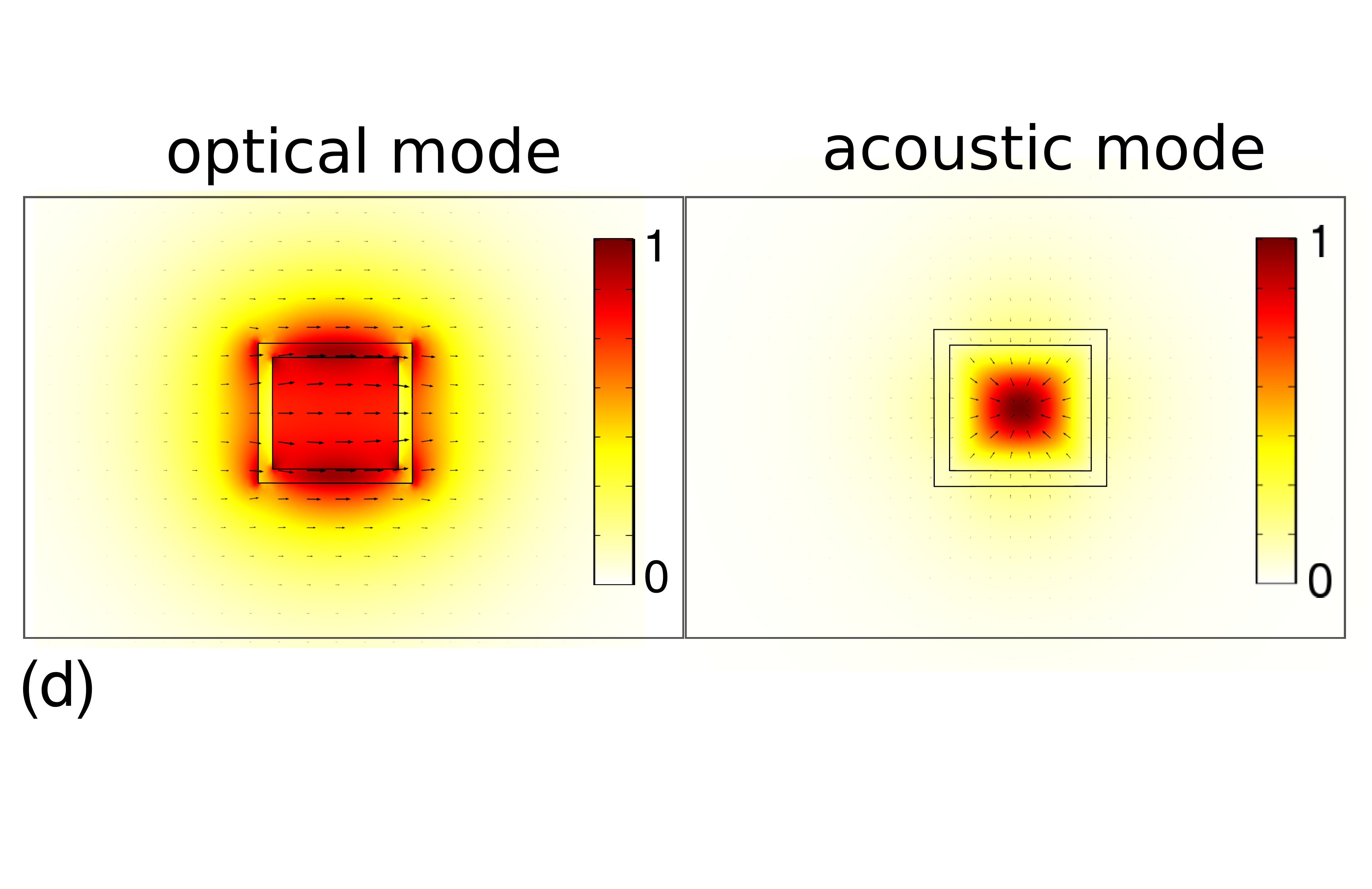} \caption{ (a) SBS power gain coefficient $g_P^\mathrm{wg}$ $\left[ \mathrm{W}^{-1} \mathrm{m}^{-1}\right]$ versus air filling fraction $f$ in the core and silicon shell thickness $d$; (b) $\log_{10}\left(Q_\mathrm{A}\right)$ where $Q_\mathrm{A}$ is acoustic quality factor; (c) Stokes shift $\Omega_\mathrm{B}^\mathrm{wg}/(2\pi)$ $\left[\mathrm{GHz} \right]$; and (d) Optical and acoustic modes at $(f=80\%,d=50~\mathrm{nm})$ where the colorscales (arb. units) refer to $\sqrt{ |\mathbf{E}_x|^2 + |\mathbf{E}_y|^2}$ and $\sqrt{|u_x|^2 + |u_y|^2}$, respectively. \label{fig:waveguidefigs} } 
\end{figure}

Figure~\ref{fig:waveguidefigs} shows results from a parameter study for our waveguide design, which has a total width $w=550~\text{nm}$ and height $h=500~\text{nm}$, and in which we vary the air filling fraction $f$ of the porous core and the thickness $d$ of the unstructured silicon. The effective material parameters from Table~1 were used for the waveguide core domain, and were interpolated where necessary. Bulk values for silicon and fused silica are given in \cite{smith2016control}. We evaluate the acousto-optic overlap as in \cite{wolff2014stimulated}, with $Q_\mathrm{A} = \Omega_\mathrm{B}^\mathrm{wg} / \Gamma_{\mathrm{B}}^\mathrm{wg}$, where the line width is given by the integral of the dynamic viscosity tensor \cite{smith2016control}. Figure~\ref{fig:waveguidefigs}a shows the gain coefficient $g_P^\mathrm{wg}$ versus filling fraction $f$ and silicon shell thickness $d$. A maximum of $g_P^\mathrm{wg} = 193 \, \mathrm{W}^{-1} \mathrm{m}^{-1}$ occurs at $(f = 80.7\%, d = 55 \, \mathrm{nm})$ which is comparable with values for chalcogenide ($\approx 300 \, \mathrm{W}^{-1} \mathrm{m}^{-1}$ \cite{pant2011chip,wolff2014germanium}) and germanium ($\approx 488 \, \mathrm{W}^{-1} \mathrm{m}^{-1}$ \cite{wolff2014germanium}) waveguides. The gain is low for small $f$ and large shell thickness due to poor acousto-optic overlap. In Fig.~\ref{fig:waveguidefigs}b we show $\log_{10}(Q_\mathrm{A})$, where the low $Q_\mathrm{A}$ region for high $f$ is due to acoustic cutoff in the waveguide. Figure~\ref{fig:waveguidefigs}c, showing the Stokes shift $\Omega_\mathrm{B}^\mathrm{wg}/ (2\pi) \, \left[ \mathrm{GHz}\right]$ for the waveguide, indicates that the large variations in $Q_\mathrm{A}$ are caused by the acoustic loss. The Stokes shift is seen to vary smoothly over the parameter space and, since, $V_\mathrm{A}^\mathrm{wg} \propto \Omega_\mathrm{B}^\mathrm{wg}$, shows that the acoustic velocity changes by as much as $33\%$. In Fig.~\ref{fig:waveguidefigs}d we present the norm of the transverse electric field $\sqrt{ |\mathbf{E}_\mathrm{x}|^2 + |\mathbf{E}_\mathrm{y}|^2}$ and the norm of the transverse acoustic field $\sqrt{|u_\mathrm{x}|^2 + |u_\mathrm{y}|^2}$ at $(f=80\%,d=50~\mathrm{nm})$, which is close to the values for the optimal gain. Here $u$ is the acoustic displacement vector and $\mathbf{E}$ denotes the electric field. These mode profiles clearly demonstrate that the optical and acoustic fields for SBS in our silicon waveguide are strongly confined. 

The electric field discontinuities at the silicon-silica and the silicon-metamaterial boundaries are small in our waveguide and so radiation pressure contributions to the SBS gain are negligible. As a result, the waveguide does not reach the very high SBS-gains that were reported in nearly or completely suspended silicon waveguides \cite{rakich2012giant,van2015interaction} which are driven primarily by radiation pressure \cite{wolff2016brillouin}. However, our structure has an advantage in that it does not need to be suspended, which makes it mechanically robust. 

\section{Discussion and Conclusions} \label{sec:discussion} We have demonstrated theoretically that the bulk SBS gain coefficient of silicon can be enhanced by two orders of magnitude using a metamaterial design based on silicon inverse opals. This enhancement is attributed to the larger compressibility of the inverse opal, which lowers the acoustic velocity, and the reduced Brillouin linewidth, which increases the SBS gain coefficient. We also designed a waveguide based on this material which demonstrates gain values comparable to chalcogenide waveguides. Further work includes developing a full numerical procedure for the shear terms of the effective photoelastic and phonon viscosity tensors, which is difficult to implement in standard FEM solvers. From a practical perspective, fabricating our metamaterial will require precision, since, for example, variations in the air hole radii and positions lead to attenuation through Rayleigh scattering. However, there have been successful attempts at creating high-quality inverse opals in silicon \cite{blanco2000large} and such techniques may be refined to create the types of waveguide structures presented here.

\section*{Acknowledgements} This work was supported by the Australian Research Council (CUDOS Centre of Excellence, CE110001018). M.L. acknowledges support from ARC Grant DP150103611.

\bibliography{SBS_meta_bib}

\end{document}